\documentclass{article}
\usepackage{amssymb}
\usepackage{amsfonts}
\usepackage{amsmath}

\begin{document}

\begin{center}
\begin{equation*}
\end{equation*}%
{\Large BCK-algebras arising from block codes}

\begin{equation*}
\end{equation*}

\bigskip Cristina FLAUT

\begin{equation*}
\end{equation*}
\end{center}

\textbf{Abstract.} {\small In this paper, we will provide an algorithm which
allows us to find a BCK-algebra starting from a given binary block
code.\bigskip }

\bigskip

\begin{equation*}
\end{equation*}

\textbf{0. Introduction}

\begin{equation*}
\end{equation*}

BCK-algebras were first introduced in mathematics in 1966 by Y. Imai and K.
Iseki, through the paper [4], as a generalization \ of the concept of
set-theoretic difference and propositional calculi. The class of \
BCK-algebras is a proper subclass of the class of BCI-algebras and there
exist several generalizations of BCK-algebras as for example: generalized
BCK-algebras [3], dual BCK-algebras [9] , BE-algebras [1], [8]. These
algebras form an important class of logical algebras and have many
applications to various domains of mathematics, such as: group theory,
functional analysis, fuzzy sets theory, probability theory, topology, etc.
For other details about BCK-algebras and about some new applications of
them, the reader is referred to [2], [5], [6], [10], [11], [12], [13] .

One of the recent applications of BCK-algebras was given in the Coding
Theory. In Coding Theory, a block code is an error-correcting code which
encode data in blocks. In the paper [7], the authors constructed a finite
binary block-codes associated to a finite BCK-algebra. At the end of the
paper, they put the question if the converse of this statement is also true.

In the present paper, we will prove that, in some circumstances, the
converse of the above statement is also true.

\begin{equation*}
\end{equation*}

\textbf{1. Preliminaries}%
\begin{equation*}
\end{equation*}

\textbf{Definition 1.1.} An algebra $(X,\ast ,\theta )$ of type $(2,0)$ is
called a \textit{BCI-algebra} if the following conditions are fulfilled:

$1)~((x\ast y)\ast (x\ast z))\ast (z\ast y)=\theta ,$ for all $x,y,z\in X;$

$2)~(x\ast (x\ast y))\ast y=\theta ,$ for all $x,y\in X;$

$3)~x\ast x=\theta ,$ for all $x\in X$;

$4)$ For all $x,y,z\in X$ such that $x\ast y=\theta ,y\ast x=\theta ,$ it
results $x=y$.

If a BCI-algebra $X$ satisfies the following identity:

$5)$ $\theta \ast x=\theta ,~$for all $x\in X,$ then $X$ is called a \textit{%
BCK-algebra}.

A BCK-algebra $X$ is called \textit{commutative }if $x\ast (x\ast y)=y\ast
(y\ast x),$ for all $x,y\in X$ and \textit{implicative }if $x\ast (y\ast
x)=x,$ for all $x,y\in X.$

The partial order relation on a BCK-algebra is defined such that $x\leq y$
if and only if $x\ast y=\theta .$

If $(X,\ast ,\theta )$ and $(Y,\circ ,\theta )$ are two BCK-algebras, a map $%
f:X\rightarrow Y$ with the property $f\left( x\ast y\right) =f\left(
x\right) \circ f\left( y\right) ,$ for all $x,y\in X,$ is called a \textit{%
BCK-algebras morphism}$.$ If $f$ is a bijective map, then $f$ is an \textit{%
isomorphism} of BCK-algebras.

In the following, we will use some notations and results given in the paper
[7] .

From now on, in whole this paper, all considered BCK-algebras are finite.

Let $A$ be a nonempty set and let $X$ \ be a BCK-algebra.\medskip

\textbf{Definition 1.2.} \ A mapping $f:A\rightarrow X$ is called a \textit{%
BCK-function} on $A.$ A \textit{cut function} \textit{of} $f$ is a map $%
f_{r}:A\rightarrow \{0,1\},r\in X,$ such that 
\begin{equation*}
f_{r}\left( x\right) =1,\text{ if and only if \ }r\ast f\left( x\right)
=\theta ,\forall x\in A.
\end{equation*}%
\newline
A \textit{cut subset} of $A$ is the following subset of $A$

\begin{equation*}
A_{r}=\{x\in A:r\ast f\left( x\right) =\theta \}.
\end{equation*}

\textbf{Remark 1.3.} Let $f:A\rightarrow X$ \ be a BCK-function on $A.$ We
define on $X$ the following binary relation 
\begin{equation*}
\forall r,s\in X,r\sim s~~\text{if~and~only~if~~}A_{r}=A_{s}.
\end{equation*}%
This relation is an equivalence relation on $X$ and we denote with $%
\widetilde{r}$ the equivalence class of the element $r\in X.\medskip $

\textbf{Remark 1.4.} ([7] ) Let $A$ be a set with $n$ elements. We consider $%
A=\{1,2,...,n\}$ and let $X$ be a BCK-algebra. For each BCK-function $%
f:A\rightarrow X,$ we can define a binary block-code of length $n.$ For this
purpose, to each equivalence class $\widetilde{x},x\in X,$ will correspond
the codeword $w_{x}=x_{1}x_{2}...x_{n}$ with $x_{i}=j,$ if and only if $%
f_{x}\left( i\right) =j,i\in A,j\in \{0,1\}.$We denote this code with $%
V_{X}. $

Let $V$ be a binary block-code and $w_{x}=x_{1}x_{2}...x_{n}\in V,$ $%
w_{y}=y_{1}y_{2}...y_{n}\in V$ ~be two codewords$.$ On $V$ we can define the
following partial order relation: 
\begin{equation}
w_{x}\preceq w_{y}\text{ if and only if }y_{i}\leq x_{i},i\in \{1,2,...,n\}.
\tag{1.1.}
\end{equation}

In the paper [7], the authors constructed binary block-codes generated \ by
BCK-functions. At the end of the paper they put the following question: 
\textit{for each binary block-code} $V$, \textit{there is a BCK-function
which determines} $V$? The answer of this question \ is partial affirmative,
as we can see in Theorem 2.2 and Theorem 2.9. 
\begin{equation*}
\end{equation*}

\textbf{2. Main results}%
\begin{equation*}
\end{equation*}

Let $(X,\leq )$ be a finite partial ordered set with the minimum element $%
\theta $ . We define the following binary relation $"\ast "~$on $X:$%
\begin{equation}
\left\{ 
\begin{array}{c}
\theta \ast x=\theta \text{ and }x\ast x=\theta ,\forall x\in X; \\ 
x\ast y=\theta ,\text{ if \ }x\leq y,\ \ \ x,y\in X; \\ 
x\ast y=x,\text{ if }y<x,~\ x,y\in X; \\ 
x\ast y=y,\text{ if }x\in X\text{ ~and~ }y\in X\text{ ~can't be compared.}%
\end{array}%
\right.  \tag{2.1.}
\end{equation}

\textbf{Proposition 2.1. }\textit{With the above notations, the algebra} $%
\left( X,\ast ,\theta \right) $ \textit{is a non-commutative and
non-implicative BCK-algebra.\medskip } $\Box \medskip $

If the above BCK-algebra has $n$ elements, we will denote it with $\mathcal{C%
}_{n}.\medskip $

Let $V$ be a binary block-code with $n$ codewords of length $n.$ We consider
the matrix $M_{V}=\left( m_{i,j}\right) _{i,j\in \{1,2,...,n\}}\in \mathcal{M%
}_{n}(\{0,1\})$ with the rows consisting of the codewords of $V.$ This
matrix is called \textit{the matrix associated to the code} $V.\medskip $

\textbf{Theorem 2.2.} \textit{With the above notations, if \ the codeword} $%
\underset{n-\text{time}}{\underbrace{11...1}}$ \textit{is in} $V$ \textit{%
and the matrix} $M_{V}$ \textit{is upper triangular with} $m_{ii}=1,$ 
\textit{for all} $i\in \{1,2,...,n\}$, \textit{there are a set} $A$ \textit{%
with} $n$ \textit{elements, a BCK-algebra }$X$ \textit{and a BCK-function} $%
f:A\rightarrow X$ \textit{such that} $f$ \textit{determines} $V.\medskip $

\textbf{Proof. }\ We consider on $V$ the lexicographic order, denoted by $%
\leq _{lex}$. \ It results that $(V,\leq _{lex})$ is a totally ordered set.
Let $V=\{w_{1},w_{2},...,w_{n}\},$ with $w_{1}\geq _{lex}w_{2}\geq
_{lex}...\geq _{lex}w_{n}.$ From here, we obtain that $w_{1}=\underset{n-%
\text{time}}{\underbrace{11...1}}$ and $w_{n}=\underset{(n-1)-\text{time}}{%
\underbrace{00...0}1}.$ On $V$ we define a partial order $\preceq $ as in
Remark 1.4. Now, $\left( V,\preceq \right) $ is a partial ordered set with $%
\ w_{1}\preceq w_{i},i\in \{1,2,...,n\}.$ We remark that $w_{1}=\theta \ $
is the "zero" in $\left( V,\preceq \right) $ and $w_{n}$ is a maximal
element in $\left( V,\preceq \right) .$ We define on $\left( V,\preceq
\right) $ a binary relation $"\ast "$ as in Proposition 2.1. It results that 
$X=\left( V,\ast ,w_{1}\right) $ becomes a BCK-algebra and $V$ is isomorphic
to $\mathcal{C}_{n}$ as \ BCK-algebras. We consider $A=V$ and the identity
map $f:A\rightarrow V,f\left( w\right) =w$ as a BCK-function. \ The
decomposition of \ $f$ provides a family of maps $V_{\mathcal{C}%
_{n}}=\{f_{r}:A\rightarrow \{0,1\}~/~$ \ $f_{r}\left( x\right) =1,$ if and
only if \ $r\ast f\left( x\right) =\theta ,\forall x\in A,r\in X\}.$ This
family is the binary block-code $V$ relative to the order relation $\preceq
. $ Indeed, let $w_{k}\in V,1<k<n,$ $w_{k}=\underset{k-1}{\underbrace{00...0}%
}x_{i_{k}}...x_{i_{n}},~~x_{i_{k}}...x_{i_{n}}\in \{0,1\}.$ If $x_{i_{j}}=0,$
it results that $w_{k}\preceq w_{i_{j}}$ and $w_{k}\ast w_{i_{j}}=\theta .$
If \ $x_{i_{j}}=1,$ we obtain that $w_{i_{j}}\preceq w_{k}$ or $w_{i_{j}}$
and $w_{k}$ can't be compared, therefore $w_{k}\ast w_{i_{j}}=w_{k}.\medskip
\Box $

\textbf{Remark 2.3.} Using technique developed in [7], we remark that a
BCK-algebra determines a unique binary block-code, but a binary block-code
as in Theorem 2.2 can be determined by two or more algebras(see Example
3.1). If two BCK-algebras, $A_{1},A_{2}$ determine the same binary
block-code, we call them \textit{code-similar algebras, } denoted by $%
A_{1}\thicksim A_{2}$. We denote by $\mathfrak{C}_{n}$ the set of the binary
block-codes of the form given in the Theorem 2.2.$\medskip $

\textbf{Remark 2.4.} If we consider $\mathfrak{B}_{n}$, the set of all
finite BCK-algebras with\textit{\ }$\ n$ elements, then the relation
code-similar is an equivalence relation on $\mathfrak{B}_{n}$\textit{.} Let $%
\mathfrak{Q}_{n}$ be the quotient set. For $V\in \mathfrak{C}_{n}$, an
equivalent class in $\mathfrak{Q}_{n}$ is $\widehat{V}=\{B\in \mathfrak{B}%
_{n}$ $/$ $B$ determines the binary block-code $V\}.\medskip $

\textbf{Proposition 2.5.} \textit{The quotient set }$\mathfrak{Q}_{n}$ 
\textit{has }$2^{\frac{\left( n-1\right) \left( n-2\right) }{2}}$ \textit{%
elements, the same cardinal as the set } $\mathfrak{C}_{n}$.\medskip

\textbf{Proof.} We will compute the cardinal \ of the set $\mathfrak{C}_{n}.$
For $V\in \mathfrak{C}_{n},$ let $M_{V}$ be its associated matrix. This
matrix is upper triangular with $m_{ii}=1,$ for all $i\in \{1,2,...,n\}.$ We
calculate in how many different ways the rows of such a matrix can be
written. The second row of the matrix $M_{V}$ has the form $\left(
0,1,a_{3},...,a_{n}\right) ,$ where $a_{3},...,a_{n}\in \{0,1\}.$ Therefore,
the number of different rows of this type is $2^{n-2}$ and it is equal with
the number of functions from a set with $n-2$ elements to the set $\{0,1\}.$
The third row of the matrix $M_{V}$ has the form $\left(
0,0,1,a_{4},...,a_{n}\right) ,$ where $a_{4},...,a_{n}\in \{0,1\}.$ In the
same way, it results that the number of different rows of this type is $%
2^{n-3}.$ Finally, we get that the cardinal of the set $\mathfrak{C}_{n}$ is 
$2^{n-2}2^{n-3}...2=$ $2^{\frac{\left( n-1\right) \left( n-2\right) }{2}}$.$%
\Box \medskip $

\textbf{Remark 2.6.} If $\mathfrak{N}_{n}$ is the number of all finite
non-isomorphic BCK-algebras with $\ n$ elements, then\textit{\ }$\mathfrak{N}%
_{n}\geq 2^{\frac{\left( n-1\right) \left( n-2\right) }{2}}.\medskip $

\textbf{Remark 2.7.} 1) Let $V_{1},V_{2}\in \mathfrak{C}_{n}$ and $%
M_{V_{1}},M_{V_{2}}$ be the associated matrices. We denote by $r_{j}^{V_{i}} 
$ a row in the \ matrix $M_{V_{i}},i\in \{1,2\},~j\in \{1,2,...,n\}.$ On $%
\mathfrak{C}_{n},~$we define the following totally order relation 
\begin{equation*}
V_{1}\succeq _{lex}V_{2}~\text{if~there~is~}i\in \{2,3,...,n\}~\text{%
such~that~}r_{1}^{V_{1}}=r_{1}^{V_{2}},...,r_{i-1}^{V_{1}}=r_{i-1}^{V_{2}}~%
\text{and~~}r_{i}^{V_{1}}\geq _{lex}r_{i}^{V_{2}},
\end{equation*}%
where $\geq _{lex}$ is the lexicographic order$.$

2) Let $V_{1},V_{2}\in \mathfrak{C}_{n}$ and $M_{V_{1}},M_{V_{2}}$ be the
associated matrices. We define a partially order on $\mathfrak{C}_{n}$ 
\begin{equation*}
V_{1}\ll V_{2}~\text{if~there~is~}i\in \{2,3,...,n\}\ \text{such\ that\ }%
r_{1}^{V_{1}}=r_{1}^{V_{2}},...,r_{i-1}^{V_{1}}=r_{i-1}^{V_{2}}\ \text{and\ }%
r_{i}^{V_{1}}\preceq r_{i}^{V_{2}},
\end{equation*}
where $\preceq $ is the order relation given by the relation $\left(
1.1\right) .$

3) Let $\Theta =\left( \theta _{ij}\right) _{i,j\in \{1,2,...,n\}}\in 
\mathcal{M}\left( \{0,1\}\right) $ be a matrix such that$~\theta _{ij}=1,$ $%
i\leq j,$ for all $i,j\in \{1,2,...,n\}$ and $\theta _{ij}=0$ in the rest.
It results that the \ code $\Omega ,$ such that $M_{\Omega }=$ $\Theta ,$ is
the minimum element in the partial ordered set $\left( \mathfrak{C}_{n},%
\text{ }\ll \right) ,$ where elements in $\mathfrak{C}_{n}$ are descending
ordered relative to $\succeq _{lex}$ defined in 1). Using the multiplication 
$"\ast "$ given in relation $\left( 2.1\right) $ and Proposition 2.1, we
obtain that $\left( \mathfrak{C}_{n},\ast ,\Omega \right) $ is a
non-commutative and non-implicative BCK-algebra\textit{.\medskip\ }Due to
the above remarks and relation $\left( 2.1\right) ,~$this BCK-algebra
determines a binary block-code $V_{\mathfrak{C}_{n}}\ $of length $2^{\frac{%
\left( n-1\right) \left( n-2\right) }{2}}.$ Obviously, $\widehat{V}_{%
\mathfrak{C}_{n}}\in \mathfrak{C}_{2^{\frac{\left( n-1\right) \left(
n-2\right) }{2}}}.\medskip $

\textbf{Proposition 2.8. }\ \textit{Let} \ $A=\left( a_{i,j}\right) 
_{\substack{ i\in \{1,2,...,n\}  \\ j\in \{1,2,...,m\}}}\in \mathcal{M}%
_{n,m}(\{0,1\})$ \textit{be a matrix with rows lexicographic ordered in the
descending sense. Starting from this matrix, we can find a \ matrix} $%
B=\left( b_{i,j}\right) _{i,j\in \{1,2,...,q\}}\in \mathcal{M}_{q}(\{0,1\}),$
$q=n+m,$ \textit{such that} $B$ \textit{is an upper triangular matrix, with }%
$b_{ii}=1,\forall i\in \{1,2,...,q\}$ \textit{and} $A$ \textit{becomes a
submatrix of the matrix} $B.\medskip $

\textbf{Proof. }We insert in the left side of the matrix $A$ ( from the
right to the left) the following $n$ new columns of the form $\underset{n}{%
\underbrace{00...01}},\underset{n}{\underbrace{00...10}},...,\underset{n}{%
\underbrace{10...00}}.$ It results a new matrix $D$ with $n$ rows and $n+m$
columns. Now, we insert in the bottom of the matrix $D$ the following $m$
rows: $\underset{n}{\underbrace{00...0}}\underset{m}{\underbrace{10...00}}$
, $\underset{n+1}{\underbrace{00...0}}\underset{m-1}{\underbrace{01...00}}%
,...,\underset{n+m-1}{\underbrace{000}}1.$ We obtained the asked matrix $%
B.\Box \medskip $

\textbf{Theorem 2.9.} \textit{With the above notations, we consider} $V$ 
\textit{\ a binary block-code with} $n$ \textit{codewords of length} $%
m,n\neq m,$ \textit{or a block-code with }$n$ \textit{codewords of length} $%
n\,\ $\textit{such that} \textit{the codeword} $\underset{n-\text{time}}{%
\underbrace{11...1}}$ \textit{is not in }$V,$\textit{\ or a block-code with }%
$n$ \textit{codewords of length} $n\,\ $\textit{such that the matrix} $M_{V}$
\textit{is not upper triangular}$.$ \textit{There are a natural number} $%
q\geq \max \{m,n\}$, \textit{a set} $A$ \textit{with} $m$ \textit{elements
and a BCK-function} $f:A\rightarrow \mathcal{C}_{q}$ \textit{such that the
obtained block-code }$V_{\mathcal{C}_{n}}$ \textit{contains the block-code} $%
V$ \textit{as a subset.\medskip }

\textbf{Proof.} Let $V$ be a binary block-code, $V=\{w_{1,}w_{2},...,w_{n}%
\}, $ with codewords of length $m.$ We consider the codewords $%
w_{1,}w_{2},...,w_{n}$ lexicographic ordered, $w_{1}\geq _{lex}w_{2}\geq
_{lex}...\geq _{lex}w_{n}.$ Let $M\in \mathcal{M}_{n,m}(\{0,1\})$ be the
associated matrix with the rows $w_{1},...,w_{n}$ in this order. Using
Proposition 2.8, we can extend the matrix $M$ to a square matrix $M^{\prime
}\in \mathcal{M}_{q}(\{0,1\}),q=m+n,$ such that $M^{\prime }=\left(
m_{i,j}^{\prime }\right) _{i,j\in \{1,2,...,q\}}$ is an upper triangular
matrix with $m_{ii}=1,$ for all $i\in \{1,2,...,q\}.$ If the first line of \
the matrix $M^{\prime }$ is not $\underset{q}{\underbrace{11...1}},$ then we
insert the row $\underset{q+1}{\underbrace{11...1}}$ as a first row and the
column $1\underset{q}{\underbrace{0...0}}~$as a first column $.$ Applying
Theorem 2.2 for the matrix $M^{\prime },$ we obtain a BCK-algebra $\mathcal{C%
}_{q}=\{x_{1},...,x_{q}\},$with $x_{1}=\theta $ the zero of the algebra $%
\mathcal{C}_{q}$ and a binary block-code $V_{\mathcal{C}_{q}}.\ $Assuming
that the initial columns of the matrix $M$ have in the new matrix $M^{\prime
}$ positions $i_{j_{1}},i_{j_{2}},...,i_{j_{m}}\in \{1,2,...,q\},$ let $%
A=\{x_{j_{1}},x_{j_{2}},...,x_{j_{m}}\}\subseteq \mathcal{C}_{q}.$ The
BCK-function $f:A\rightarrow \mathcal{C}_{q},f\left( x_{j_{i}}\right) =$ $%
x_{j_{i}},$ $i\in \{1,2,...,m\},$ determines the binary block-code $V_{%
\mathcal{C}_{q}}$ such that $V\subseteq V_{\mathcal{C}_{q}}.\Box $

\begin{equation*}
\end{equation*}

\bigskip \textbf{3. Examples}

\begin{equation*}
\end{equation*}

\textbf{Example 3.1. }\ Let $V=\{0110,0010,1111,0001\}$ be a binary block
code. Using the lexicographic order, the code $V$ can be written \newline
$V=\{1111,0110,0010,0001\}=\{w_{1},w_{2},w_{3},w_{4}\}.$ From Theorem 2.2,
defining the partial order $\preceq $ on $V,$ we remark that $w_{1}\preceq
w_{i},i\in \{2,3,4\},w_{2}\preceq w_{3},w_{2}$ can't be compared with $w_{4}$
and $w_{3}$ can't be compared with $w_{4}$. The operation $"\ast "$ on $V$
is given in the following table:

\begin{tabular}{l|llll}
$\ast $ & $w_{1}$ & $w_{2}$ & $w_{3}$ & $w_{4}$ \\ \hline
$w_{1}$ & $w_{1}$ & $w_{1}$ & $w_{1}$ & $w_{1}$ \\ 
$w_{2}$ & $w_{2}$ & $w_{1}$ & $w_{1}$ & $w_{2}$ \\ 
$w_{3}$ & $w_{3}$ & $w_{3}$ & $w_{1}$ & $w_{3}$ \\ 
$w_{4}$ & $w_{4}$ & $w_{4}$ & $w_{4}$ & $w_{1}$%
\end{tabular}%
.

Obviously, $V$ with the operation $"\ast "$ is a BCK-algebra.

We remark that the same binary block code $V$ can be obtained from the
BCK-algebra $(A,\circ ,\theta )$

\begin{tabular}{l|llll}
$\circ $ & $\theta $ & $a$ & $b$ & $c$ \\ \hline
$\theta $ & $\theta $ & $\theta $ & $\theta $ & $\theta $ \\ 
$a$ & $a$ & $\theta $ & $\theta $ & $a$ \\ 
$b$ & $b$ & $a$ & $\theta $ & $b$ \\ 
$c$ & $c$ & $c$ & $c$ & $\theta $%
\end{tabular}%
\newline
with BCK-function, $f:V\rightarrow V,f(x)=x.$(see [7] , Example 4.2). From
the associated Cayley multiplication tables, it is obvious that the algebras 
$(A,\circ ,\theta )$ and $(V,\ast ,w_{1})$ are not isomorphic. From here, we
obtain that BCK-algebra associated to a binary block-code as in Theorem 2.2
is not unique up to an isomorphism. We remark that the BCK-algebra $(A,\circ
,\theta )$ is commutative and non implicative and BCK-algebra $(V,\ast
,w_{1})$ is non commutative and non implicative. Therefore, if we start from
commutative BCK-algebra $(A,\circ ,\theta )$ to obtain the code $V,$ as in
[7], and then we construct the BCK-algebra $(V,\ast ,w_{1}),$ as in Theorem
2.2, the last obtained algebra lost the commutative property even that these
two algebras are code-similar.\medskip

\textbf{Example 3.2.} \ Let $X$ be a non empty set and $\mathfrak{F}%
=\{f:X\rightarrow \{0,1\}~/$ $f$ function$\}.$ On $\mathfrak{F}$ is defined
the following multiplication%
\begin{equation*}
(f\circ g)\left( x\right) =f\left( x\right) -\text{\textit{min}}\{f\left(
x\right) ,g\left( x\right) \},\forall x\in X.
\end{equation*}

$\left( \mathfrak{F},\circ ,\mathbf{0}\right) $, where $\mathbf{0}\left(
x\right) =0,\forall x\in X,$ is an implicative BCK-algebra([12], Theorem 3.3
and Example 1).

If $X$ is a set with three elements, we can consider $\mathfrak{F}%
=\{000,001,010,011,100,101,110,111\}$ the set of binary block-codes of
length $3.$ We \ have the following multiplication table.

\begin{tabular}{l|l|l|l|l|l|l|l|l|l|}
\cline{2-10}
$\circ $ & $000$ & $001$ & $010$ & $011$ & $100$ & $101$ & $110$ & $111$ & 
{\tiny The obtained binary code-words} \\ \hline
\multicolumn{1}{|l|}{$000$} & $000$ & 000 & 000 & 000 & 000 & 000 & 000 & 000
& 11111111 \\ \hline
\multicolumn{1}{|l|}{$001$} & 001 & 000 & 001 & 000 & 001 & 000 & 001 & 000
& 01010101 \\ \hline
\multicolumn{1}{|l|}{$010$} & 010 & 010 & 000 & 000 & 010 & 010 & 000 & 000
& 00110011 \\ \hline
\multicolumn{1}{|l|}{$011$} & 011 & 010 & 001 & 000 & 011 & 010 & 001 & 000
& 00010001 \\ \hline
\multicolumn{1}{|l|}{$100$} & 100 & 100 & 100 & 100 & 000 & 000 & 000 & 000
& 00001111 \\ \hline
\multicolumn{1}{|l|}{$101$} & 101 & 100 & 101 & 100 & 001 & 000 & 001 & 000
& 00000101 \\ \hline
\multicolumn{1}{|l|}{$110$} & 110 & 110 & 100 & 100 & 010 & 010 & 000 & 000
& 00000011 \\ \hline
\multicolumn{1}{|l|}{$111$} & 111 & 110 & 101 & 100 & 011 & 010 & 001 & 000
& 00000001 \\ \hline
\end{tabular}%
.

\bigskip

We obtain the following binary block-code\newline
$V=\{11111111,01010101,00110011,00010001,$\newline
$00001111,00000101,00000011,00000001\},$ with the elements lexicographic
ordered in the descending sense$.$ From Theorem 2.2, defining the partial
order $\preceq $ on $V$ and the multiplication $"\ast ",$ we have that $%
\left( V,\ast ,11111111\right) $ is a non-implicative BCK-algebra and the
algebras $\left( V,\ast ,11111111\right) $ and $\left( \mathfrak{F},\circ ,%
\mathbf{0}\right) $ are code-similar.\medskip

\textbf{Example 3.3. }Let $V=\{11110,10010,10011,00000\}$ be a binary block
code. Using the lexicographic order, the code $V$ can be written \newline
$V=\{11110,10011,10010,00000\}=\{w_{1},w_{2},w_{3},w_{4}\}.$ Let $M_{V}\in 
\mathcal{M}_{4,5}\left( \{0,1\}\right) $ be the associated matrix$,$ $%
M_{V}=\left( 
\begin{tabular}{lllll}
$1$ & $1$ & $1$ & $1$ & $0$ \\ 
$1$ & $0$ & $0$ & $1$ & $1$ \\ 
$1$ & $0$ & $0$ & $1$ & $0$ \\ 
$0$ & $0$ & $0$ & $0$ & $0$%
\end{tabular}%
\right) .$ Using Proposition 2.8, we construct an upper triangular matrix,
starting from the matrix $M_{V}.$ It \ results the following matrices: 
\newline
$D=\left( 
\begin{tabular}{lllllllll}
$1$ & $0$ & $0$ & $0$ & $\mathbf{1}$ & $\mathbf{1}$ & $\mathbf{1}$ & $%
\mathbf{1}$ & $\mathbf{0}$ \\ 
$0$ & $1$ & $0$ & $0$ & $\mathbf{1}$ & $\mathbf{0}$ & $\mathbf{0}$ & $%
\mathbf{1}$ & $\mathbf{1}$ \\ 
$0$ & $0$ & $1$ & $0$ & $\mathbf{1}$ & $\mathbf{0}$ & $\mathbf{0}$ & $%
\mathbf{1}$ & $\mathbf{0}$ \\ 
$0$ & $0$ & $0$ & $1$ & $\mathbf{0}$ & $\mathbf{0}$ & $\mathbf{0}$ & $%
\mathbf{0}$ & $\mathbf{0}$%
\end{tabular}%
\right) $ and \newline
$B=\left( 
\begin{array}{ccccccccc}
1 & 0 & 0 & 0 & \mathbf{1} & \mathbf{1} & \mathbf{1} & \mathbf{1} & \mathbf{0%
} \\ 
0 & 1 & 0 & 0 & \mathbf{1} & \mathbf{0} & \mathbf{0} & \mathbf{1} & \mathbf{1%
} \\ 
0 & 0 & 1 & 0 & \mathbf{1} & \mathbf{0} & \mathbf{0} & \mathbf{1} & \mathbf{0%
} \\ 
0 & 0 & 0 & 1 & \mathbf{0} & \mathbf{0} & \mathbf{0} & \mathbf{0} & \mathbf{0%
} \\ 
0 & 0 & 0 & 0 & 1 & 0 & 0 & 0 & 0 \\ 
0 & 0 & 0 & 0 & 0 & 1 & 0 & 0 & 0 \\ 
0 & 0 & 0 & 0 & 0 & 0 & 1 & 0 & 0 \\ 
0 & 0 & 0 & 0 & 0 & 0 & 0 & 1 & 0 \\ 
0 & 0 & 0 & 0 & 0 & 0 & 0 & 0 & 1%
\end{array}%
\right) .$

Since the first row is not $\underset{9}{\underbrace{11...1}},$ using
Theorem 2.8, we insert a new row $\underset{10}{\underbrace{11...1}}$ as a
first row and a new column $\underset{10}{\underbrace{10...0}}$ as a first
column. We obtain the following matrix: $B^{\prime }=\left( 
\begin{array}{cccccccccc}
1 & 1 & 1 & 1 & 1 & \mathbf{1} & \mathbf{1} & \mathbf{1} & \mathbf{1} & 
\mathbf{1} \\ 
0 & 1 & 0 & 0 & 0 & \mathbf{1} & \mathbf{1} & \mathbf{1} & \mathbf{1} & 
\mathbf{0} \\ 
0 & 0 & 1 & 0 & 0 & \mathbf{1} & \mathbf{0} & \mathbf{0} & \mathbf{1} & 
\mathbf{1} \\ 
0 & 0 & 0 & 1 & 0 & \mathbf{1} & \mathbf{0} & \mathbf{0} & \mathbf{1} & 
\mathbf{0} \\ 
0 & 0 & 0 & 0 & 1 & \mathbf{0} & \mathbf{0} & \mathbf{0} & \mathbf{0} & 
\mathbf{0} \\ 
0 & 0 & 0 & 0 & 0 & 1 & 0 & 0 & 0 & 0 \\ 
0 & 0 & 0 & 0 & 0 & 0 & 1 & 0 & 0 & 0 \\ 
0 & 0 & 0 & 0 & 0 & 0 & 0 & 1 & 0 & 0 \\ 
0 & 0 & 0 & 0 & 0 & 0 & 0 & 0 & 1 & 0 \\ 
0 & 0 & 0 & 0 & 0 & 0 & 0 & 0 & 0 & 1%
\end{array}%
\right) .$\newline
The binary block-code $W=\{w_{1},...,w_{10}\},$ whose codewords are the rows
of the matrix $B^{\prime },$ determines a BCK-algebra $(X,\ast ,w_{1}).$ Let 
$A=\{w_{6},w_{7},w_{8},w_{9},w_{10}\}$ and $f:A\rightarrow X,f\left(
w_{i}\right) =w_{i},i\in \{6,7,8,9,10\}$ be a BCK-function which determines
the binary block-code\newline
$U=\{11111,11110,10011,10010,00000,10000,01000,00100,00010,00001\}.$ The
code $V$ is a subset of the code $U.\medskip $

\textbf{Conclusions.} In this paper, we proved that to each binary
block-code $V$ we can associate a BCK-algebra $X$ such that the binary
block-code generated by $X,V_{X},$ contains the code $V$ as a subset. In
some particular case, we have $V_{X}=V.$

From Example 3.1 and 3.2, we remark that two code-similar BCK-algebras can't
have the same properties. For example, some algebras from the same
equivalence class can be commutative and other non-commutative or some
algebras from the same equivalence class can be implicative and other
non-implicative. As a further research, will be very interesting to study
what common properties can have two code-similar BCK-algebras.

Due to this connection of BCK-algebras with Coding Theory, we can consider
the above results as a starting point in the study of new applications of
these algebras in the Coding Theory. 
\begin{equation*}
\end{equation*}%
\textbf{Acknowledgements}

I want thank to anonymous referees for their comments, suggestions and ideas
which helped me to improve this paper. The author also thanks Professor
Arsham Borumand Saeid for having brought [7] to my attention.%
\begin{equation*}
\end{equation*}%
\textbf{References}%
\begin{equation*}
\end{equation*}

[1] S. Abdullah, A. F. Ali, JIFS, \textit{Applications of N-structures in
implicative filters of BE-algebras}, will appear in J. Intell. Fuzzy Syst.,
DOI 10.3233/IFS-141301.

[2] J. S. Han, H. S. Kim, J. Neggers, \textit{On linear fuzzifications of
groupoids with special emphasis on BCK-algebras}, J. Intell. Fuzzy Syst., 
\textbf{24(1)(2013)}, 105-110.

[3] S. M. Hong, Y. B. Jun, M. A. \"{O}zt\"{u}rk, \textit{Generalizations of
BCK-algebras}, Sci. Math. Jpn. Online, \textbf{8(2003)}, 549--557

[4] Y. Imai, K. Iseki, \textit{On axiom systems of propositional calculi},
Proc. Japan Academic, \textbf{42(1966)}, 19-22.

[5] K. Is\'{e}ki, S. Tanaka, \textit{An introduction to the theory of
BCK-algebras}, Math. Jpn. \textbf{23(1978)}, 1--26.

[6]Young Bae Jun, \textit{Soft BCK/BCI-algebras}, Comput. Math. Appl., 
\textbf{56(2008)}, 1408--1413.

[7] Y. B. Jun, S. Z. Song, \textit{Codes based on BCK-algebras}, Inform.
Sciences., \textbf{181(2011)}, 5102-5109.

[8] H. S. Kim and Y. H. Kim, \textit{On BE-algebras}, Sci. Math. Jpn. Online
e-2006\textbf{(2006)}, 1199-1202.

[9] K. H. Kim, Y. H. Yon, \textit{Dual BCK-algebra and MV -algebra}, Sci.
Math. Jpn., \textbf{66(2007)}, 247-253.

[10] A.B. Saeid, \textit{Redefined fuzzy subalgebra (with thresholds) of\
BCK/BCI-algebras,} Iran. J. Math. Sci. Inform, \textbf{4(2)(2009)}, 9-24.

[11] A. B. Saeid, M. K. Rafsanjani, D. R. Prince Williams, \textit{Another
Generalization of Fuzzy BCK/BCI-Algebras}, Int. J. Fuzzy Syst., \textbf{%
14(1)(2012)}, 175-184.

[12] Z. Samaei, M. A. Azadani, L. Ranjbar, \textit{A Class of BCK-Algebras},
Int. J. Algebra, \textbf{5(28)(2011)}, 1379 - 1385.

[13] X. Xin, Y. Fu, \textit{Some results of convex fuzzy sublattices}, J.
Intell. Fuzzy Syst., \textbf{27(1)(2014)}, 287-298.%
\begin{equation*}
\end{equation*}%
Cristina FLAUT

{\small Faculty of Mathematics and Computer Science, Ovidius University,}

{\small Bd. Mamaia 124, 900527, CONSTANTA, ROMANIA}

{\small http://cristinaflaut.wikispaces.com/;
http://www.univ-ovidius.ro/math/}

{\small e-mail: cflaut@univ-ovidius.ro; cristina\_flaut@yahoo.com}

\end{document}